\documentclass[aps,twocolumn,pra,superscriptaddress,amsmath,showpacs,tightenlines,pdflatex]{revtex4}

\usepackage{amsmath}

\usepackage{graphicx,times}
\usepackage{amstext}
\usepackage{amsmath}            %serve per le subequazioni
\usepackage{amssymb}            %serve per il simbolo "marchio registrato", \circledR
\usepackage{graphicx}           %serve per le figure eps, ps etcyy
\usepackage{latexsym}
\usepackage{bm}
 \usepackage{color}

 \def\Title#1{\emph{#1}}

\def \etal{\textit{et al.}}

 \def\fig#1{{#1}}
\def\tr{{\rm Tr}}

\def\CP{{\rm CP}}
\def\NP{{\rm NP}}
\def\REEP{{\rm REEP}}

\def\<{\langle}
\def\>{\rangle}

\newcommand{\RHO}[1]{{\rho_{\rm #1}}}     % 2-qubit rho
\newcommand{\SIGMA}[1]{{\sigma_{\rm #1}}} % 1-qubit rho

\newcommand{\ket}[1]{\mbox{$|#1\rangle$}}
\newcommand{\bra}[1]{\mbox{$\langle#1|$}}

\begin{document}

\title{Increasing relative nonclassicality quantified by standard entanglement potentials
by dissipation and unbalanced beam splitting}
\author{Adam Miranowicz}
\affiliation{CEMS, RIKEN, 351-0198 Wako-shi, Japan}
\affiliation{Faculty of Physics, Adam Mickiewicz University,
61-614 Pozna\'n, Poland}

\author{Karol Bartkiewicz}
\affiliation{Faculty of Physics, Adam Mickiewicz University,
61-614 Pozna\'n, Poland}\affiliation{RCPTM, Joint Laboratory of
Optics of Palack\'y University and Institute of Physics of AS CR,
Palack\'y University, 17. listopadu 12, 771 46 Olomouc, Czech
Republic}

\author{Neill Lambert}
\affiliation{CEMS, RIKEN, 351-0198 Wako-shi, Japan}

\author{Yueh-Nan Chen}
\affiliation{Department of Physics and National Center for
Theoretical Sciences, National Cheng-Kung University, Tainan 701,
Taiwan} \affiliation{CEMS, RIKEN, 351-0198 Wako-shi, Japan}

\author{Franco Nori}
\affiliation{CEMS, RIKEN, 351-0198 Wako-shi, Japan}
\affiliation{Department of Physics, The University of Michigan,
Ann Arbor, MI 48109-1040, USA}

\date{\today}

\begin{abstract}
If a single-mode nonclassical light is combined with the vacuum on
a beam splitter, then the output state is entangled. As proposed
in [Phys. Rev. Lett. \textbf{94}, 173602 (2005)], by measuring
this output-state entanglement for a balanced lossless beam
splitter, one can quantify the input-state nonclassicality. These
measures of nonclassicality (referred to as entanglement
potentials) can be based, in principle, on various entanglement
measures, leading to the negativity (NP) and concurrence (CP)
potentials, and the potential for the relative entropy of
entanglement (REEP). We search for the maximal relative
nonclassicality, which can be achieved by comparing two
entanglement measures for (i) arbitrary two-qubit states and (ii)
those which can be generated from a photon-number qubit via a
balanced lossless beam splitter, where the qubit basis states are
the vacuum and single-photon states. Surprisingly, we find that
the maximal relative nonclassicality, measured by the REEP for a
given value of the NP, can be increased (if NP$<$0.527) by using
either a tunable beam splitter or by amplitude damping of the
output state of the balanced beam splitter. We also show that the
maximal relative nonclassicality, measured by the NP for a given
value of the REEP, can be increased by phase damping (dephasing).
Note that the entanglement itself is not increased by these losses
(since they act locally), but the possible ratios of different
measures are affected. Moreover, we show that partially-dephased
states can be more nonclassical than both pure states and
completely-dephased states, by comparing the NP for a given value
of the REEP. Thus, one can conclude that not all standard
entanglement measures can be used as entanglement potentials.
Alternatively, one can infer that a single balanced lossless beam
splitter is not always transferring the whole nonclassicality of
its input state into the entanglement of its output modes. The
application of a lossy beam splitter can solve this problem at
least for the cases analyzed in this paper.
\end{abstract}


% 42.50.Xa Optical tests of quantum theory
% 03.67.Mn Entanglement measures, witnesses, and other characterizations
% 03.67.Bg Entanglement production and manipulation
% 42.50.Ex Optical implementations of quantum information processing and transfer

\maketitle

%------------------------------------------------------------------
\section{Introduction}

Nonclassical light plays a central role in quantum
optics~\cite{VogelBook,PerinaBook} and atom
optics~\cite{HarocheBook} leading to various applications in
quantum technologies, including quantum cryptography and
communication, and optical quantum information
processing~\cite{KokBook}.

In a sense, all states of light are quantum, so which of them can
be considered classical? It is usually assumed that coherent
states are classical. Thus, also their mixtures (e.g., thermal
states) are classical. All other states of light are considered
nonclassical (or quantum). Formally, this criterion can be given
in terms of the Glauber-Sudarshan
$P$~function~\cite{Glauber63,Sudarshan63}: A given state $\sigma$
is nonclassical if and only if it is not described by a positive
(semidefinite) function $P(\sigma)$. Thus, all finite
superpositions (except for the vacuum) of arbitrary Fock states
are nonclassical. We note that this definition ``hides some
serious problems'', as discussed in, e.g., Ref.~\cite{Wunsche04}.

Various methods and criteria have been devised to test whether a
given state of light is nonclassical (see, e.g.,
Refs.~\cite{VogelBook,DodonovBook,PerinaBook,Miran10,Bartkowiak11}).
However, with respect to the above definition, it seems much more
interesting physically to quantify the nonclassicality of light
rather than only to test (detect) it. For the last thirty years
various measures of nonclassicality have been studied (for reviews
see Refs.~\cite{DodonovBook,VogelBook}). The most popular of them
include entanglement potentials, nonclassical
distance~\cite{Hillery87}, and nonclassical
depth~\cite{Lee91,Lutkenhaus95}. For a comparative studies of
these measures see recent Refs.~\cite{Miran15,Arkhipov15} and
references therein. Other measures or parameters of
nonclassicality were described in, e.g., the recent
Refs.~\cite{Mari11,Gehrke12,Nakano13,Meznaric13}. Many studies
have been devoted to the nonclassical volume~\cite{Kenfack04},
which is the volume of the negative part of the Wigner function of
a given state. But it should be stressed that this nonclassical
volume is not a good measure but only a parameter of
nonclassicality, as it vanishes for some nonclassical states,
including ideal squeezed states.

In this paper we solely analyze universal nonclassicality measures
defined via entanglement potentials, which are closely related to
standard entanglement measures. Specifically, as introduced by
Asboth \etal~\cite{Asboth05}, one can quantify the nonclassicality
of a given single-mode state by measuring the entanglement
generated from this state and the vacuum by an auxiliary balanced
beam splitter (BS). This approach is operationally much simpler
than other nonclassicality measures, including the mentioned
nonclassical depth and distance.

To be more specific, the nonclassicality of a single-mode state
$\sigma$ can be quantified, according to Ref.~\cite{Asboth05}, by
the entanglement of the output state $\rho_{\rm out}$ of an
auxiliary lossless balanced BS with the state $\sigma$ and the
vacuum $\ket{0}$ at the inputs, i.e.,
\begin{equation}
  \rho_{\rm out} = U_{\rm BS} (\sigma\otimes \ket{0}\bra{0} )U^\dagger_{\rm
  BS},
\label{RhoOut1}
\end{equation}
where $U_{\rm BS}=\exp(-iH\theta)$, with $\hbar=1$ and
$\theta=\pi/2$, is the unitary transformation for a balanced BS,
which can be given in terms of the Hamiltonian
$H=\frac12i(a_{1}^\dagger a_{2}-a_{1} a_{2}^\dagger),$ where
$a_{1,2}$ ($a_{1,2}^\dagger$) are the annihilation (creation)
operators of the input modes. This transformation $U_{\rm BS}$ is
equivalent to that applied in Ref.~\cite{Miran15} up to a local
unitary transformation, which does not change the entanglement of
$\rho_{\rm out}$, as quantified by any ``good'' entanglement
measure. Note that linear transformations (like that performed by
a BS) do not change the nonclassicality of a given state. The
output state $\rho_{\rm out}$ is entangled if and only if the
input state $\sigma$ is nonclassical. In particular, for the input
in an arbitrary finite-dimensional state, except for the vacuum
state, the output state is entangled. It should be stressed that
the standard entanglement potentials, as proposed (and numerically
verified) in Ref.~\cite{Asboth05}, are based solely on the special
case of $\rho_{\rm out}$ for $\theta=\pi/2$, i.e., when the BS
transmissivity is equal to 1/2 corresponding to a balanced (50/50)
BS.

In this paper, we compare three standard entanglement potentials
of a single-qubit input state $\sigma$ corresponding to the
entanglement measures of the two-qubit output state $\rho_{\rm
out}$. The analyzed qubit is assumed to be in an arbitrary
(coherent or incoherent) superposition of the vacuum and
single-photon Fock states, so it can be referred to as a
photon-number qubit. We quantify the nonclassicality of $\sigma$
by the negativity potential (NP), concurrence potential (CP), and
the potential for the relative entropy of entanglement (REEP):
\begin{eqnarray}
  \NP(\sigma) &=& N (\rho_{\rm out}), \label{NP} \\
  \CP(\sigma) &=& C (\rho_{\rm out}), \label{CP} \\
  \REEP(\sigma) &=& E_R (\rho_{\rm out}), \label{REEP}
\end{eqnarray}
which are defined via the negativity $N$, concurrence $C$, and REE
$E_R$ for $\rho_{\rm out}$. Although these entanglement measures
are well-known, for clarity, we give their definitions and
operational interpretations in Appendix~A. The REEP is also
referred to as the entropic entanglement potential in the original
Ref.~\cite{Asboth05}. We emphasize again that we refer to
\emph{entanglement potential} of a given state $\sigma$ for
\emph{any} entanglement measure applied to the output $\rho_{\rm
out}$.

Various entanglement potentials, which are based on this unified
approach of measuring nonclassicality and entanglement, have been
attracting increasing interest especially during the past year
(see, e.g.,
Refs.~\cite{Vogel14,Mraz14,Miran15,Tasgin15,Killoran15,Ge15}). The
ultimate goal of such studies is to demonstrate how
nonclassicality can  operationally be used as a resource for
quantum technologies and quantum information processing. Thus, one
of the most natural and fundamental questions in this area
addresses the relationship between entanglement and other types of
nonclassicality. For example, the problem of faithfully converting
single-system nonclassicality into entanglement in general
mathematical terms was addressed in Ref.~\cite{Killoran15} with
the following conclusions: ``These results generalize and link
convertibility properties from the resource theory of coherence,
spin coherent states, and optical coherent states, while also
revealing important connections between local and non-local
quantum correlations.'' A conservation relation of nonclassicality
and entanglement in a balanced beam splitter was found for some
limited classes of Gaussian states in Ref.~\cite{Ge15}. A related
single-mode nonclassicality measure based on the
Simon-Peres-Horodecki criterion was described for Gaussian states
in Ref.~\cite{Tasgin15}. Refs.~\cite{Vogel14,Mraz14} demonstrated
that the rank of the two-mode entanglement of a single-mode state
$\sigma$ is equal to the rank of the expansion of $\sigma$ in
terms of classical coherent states. It was shown in
Ref.~\cite{Miran15}  that the CP of a single qubit state $\sigma$
can be interpreted as a Hillery-type nonclassical distance,
defined by the Bures distance of $\sigma$ to the vacuum, which is
the closest classical state in the single-qubit Hilbert space.
Moreover, it is known that the statistical mixtures of the vacuum
and single-photon states can be more nonclassical than their
superpositions, as shown in Ref.~\cite{Miran15} by comparing the
NP and CP.

Here we present a comparative study of these three entanglement
potentials, to show that such quantified relative nonclassicality,
i.e., the nonclassicality of one measure relative to another, can
be increased by damping and unbalanced beam splitting. Moreover,
in Ref.~\cite{Miran15} we showed that both pure and
completely-dephased single-qubit states can be considered the most
nonclassical by comparing some nonclassicality measures. Here we
show that also some partially-dephased states can be the most
relatively nonclassical in terms of the highest negativity
potential for a given value of the REE potential.

The paper is organized as follows. In Sec.~II, we calculate the
entanglement potentials for single-qubit states.  In Sec.~III, we
find the most nonclassical single-qubit states via the standard
entanglement potentials. In this section and also in Sec.~IV, we
show that there are two-qubit states, which are more entangled
than those which can be generated from a single-qubit state and
the vacuum by a balanced lossless BS. In Sec.~V, we show that by
applying a tunable and/or lossy BS, we can generate  relative
entanglement higher than that in the standard approach. We refer
to this modified approach as being based on generalized
entanglement potentials. Moreover, for the completeness and
clarity of our presentation, we recall known results in
appendices, including definitions of the standard entanglement
measures and boundary states for arbitrary two-qubit states. We
conclude in Sec.~VI.

%------------------------------------------------------------------
\section{Entanglement potentials for single-qubit states}

To make our presentation simple and convincing, we analyze, as in
Ref.~\cite{Miran15}, the nonclassicality of only single-qubit
states:
\begin{equation}
\sigma(p,x) = \sum_{m,n=0}^1\sigma_{mn} \ket{m} \bra{n} =
\left[\begin{array}{cc}
1-p&x\\
x^*&p
\end{array}\right],
 \label{rho}
\end{equation}
which are spanned by the vacuum $\ket{0}$ and the single-photon
Fock state $\ket{1}$. Here $p\in[0,1]$ is the mixing parameter,
and $|x|\in[0,\sqrt{p(1-p)}]$, which is often interpreted as a
coherence parameter. Analogously, in the context of the NMR
spectroscopy of a spin qubit, $x$ and $x^*$ are called the
\emph{coherences} between the states $\ket{0}$ and
$\ket{1}$~\cite{LevittBook}. The relation between the coherence
and entanglement of a partially coherent state of light was
recently studied in Ref.~\cite{Svozilik15}. An experimental
demonstration of the nonclassicality of the optical qubit states,
given in Eq.~(\ref{rho}), was reported in Ref.~\cite{Lvovsky02}
based on the nonclassicality criterion of Vogel~\cite{VogelBook}.
Note that the only classical state of $\sigma(p,x)$ is for $p=0$,
corresponding to the vacuum. Equation~(\ref{RhoOut1}) for
$\sigma(p,x)$ simplifies to
\begin{equation}
  \rho_{\rm out}(p,x) =\left[\begin{array}{cccc}
1-p& -\frac1{\sqrt{2}}x &\frac1{\sqrt{2}}x& 0\\
-\frac1{\sqrt{2}}x^*&\frac12 p&-\frac12p&0\\
\frac1{\sqrt{2}}x^*& -\frac12 p&\frac12p&0\\
0&0&0&0\\
\end{array}\right].
\label{RhoOut2}
\end{equation}
Here we study how well the entanglement potentials can serve as
measures of nonclassicality for a balanced BS.

As shown in Ref.~\cite{Miran15}, the concurrence potential is
given by a simple formula
\begin{equation}
  \CP[\sigma(p,x)]=p,
 \label{conc}
\end{equation}
for the arbitrary state given in Eq.~(\ref{rho}). Note that this
potential is independent of the coherence parameter $x$.
Surprisingly, the negativity potential is given by a much more
complicated formula
\begin{equation}
  \NP[\sigma(p,x)]=\frac{1}{3} \left[2 {\rm Re}
  \left(\sqrt[3]{2 \sqrt{\alpha_1}+2 \alpha_2}\right)+p-2\right],
 \label{NPgeneral}
\end{equation}
for a general state $\sigma(p,x)$, where $\alpha_1$ ($\alpha_2$)
is a polynomial of the 6th (3rd) order in $p$ and the 6th (2nd)
order in $|x|$, as explicitly given in Ref.~\cite{Miran15}.
Obviously, Eq.~(\ref{NPgeneral}) simplifies considerably for
special states, including those studied in the next sections.
Equation~(\ref{NPgeneral}) can be obtained from the formula valid
for an arbitrary two-qubit state $\rho$~\cite{Bartkiewicz15}:
\begin{equation}
48D + 3N^4 + 6N^3 - 6N^2\Pi'_2  %&&\nonumber\\
- 4N(3\Pi'_2 - 2\Pi'_3 )  = 0, \label{Negativity2}
\end{equation}
expressing the negativity via the invariant moments $\Pi'_n
=\Pi_n-1= \tr[(\rho^\Gamma)^n]-1$ and the determinant
$D=\det\rho^\Gamma$, where $\rho^\Gamma$ denotes the partially
transposed $\rho$. These moments are directly measurable, as shown
in Refs.~\cite{Bartkiewicz15b}. This general formula simplifies
for the special two-qubit states $\rho=\rho_{\rm out}$, which are
generated by a balanced BS. Thus, we can simply express the
coherence parameter $|x|$ as a function of the negativity (or
negativity potential) and the mixing parameter $p$ as follows:
\begin{equation}
  |x| = f(p,N)= \tfrac12 \sqrt{(1+p/N) [2N(N+1)-(N+p)^2]}
 \label{f}
\end{equation}
for  any $p\in[N,\sqrt{2N(N+1)}-N]$. Thus, an arbitrary
single qubit state $\sigma(p,x)$ can be given as
\begin{eqnarray}
  \sigma'(p,N,\phi) &\equiv& \sigma[p,x=f(p,N) \exp(i\phi)],
\label{rho_pN}
\end{eqnarray}
where $\phi={\rm Arg}(x)$ is the phase factor of the coherence
parameter $x$. In the context of our nonclassicality analysis, the
inclusion of $\phi$ in Eqs.~(\ref{rho}) and~(\ref{rho_pN}) is
actually irrelevant, as any ``good'' nonclassicality measures
(including the entanglement potentials) do not depend on $\phi$.
Thus, for simplicity, we can set $\phi=0$.

The calculation of the REE potential is even more demanding as
explained in Appendix~A. We have calculated the REE analytically
only for some special states, including pure and
completely-dephased single-qubit states, and the
completely-dephased output states of a BS. In other cases, the REE
potential is calculated only numerically based on semidefinite
algorithm implemented in Ref.~\cite{Girard15}.

%------------------------------------------------------------------
\section{Most relatively nonclassical single-qubit states via entanglement potentials}

Here we address the following question: which single-qubit states
are the most nonclassical if quantified by one entanglement
potential relative to another. Specifically, we compare the nonclassicality of states
for a given entanglement potential assuming that the states have
the same nonclassicality in terms of another entanglement
potential.

Figure~1 shows such comparison of the three entanglement
potentials for randomly generated single-qubit states $\sigma$.
These graphs are obtained as follows: For a given $\sigma$, we
calculated the potentials $\NP(\sigma)$, $\CP(\sigma)$, and
$\REEP(\sigma)$, and then plotted a point at
$[\NP(\sigma),\CP(\sigma)]$ in Fig. 1(a), another point at
$[\REEP(\sigma),\CP(\sigma)]$ in Fig. 1(b), and
$[\REEP(\sigma),\NP(\sigma)]$ in Fig. 1(c). The Monte-Carlo
simulated points occupy limited areas. Now we discuss the states
on their boundaries. Note that the red curves in Fig.~1 show the
boundaries of the entanglement of arbitrary two-qubit states
$\rho$ (see Appendix~B), instead of $\rho_{\rm out}$ only.
Figure~2 shows more explicitly that there is a two-qubit
entanglement (corresponding to the red regions), which cannot be
generated from single-qubit states and the vacuum by a balanced
lossless BS.

%------------------------------------------------------------------
\begin{figure}
\includegraphics[width=6cm]{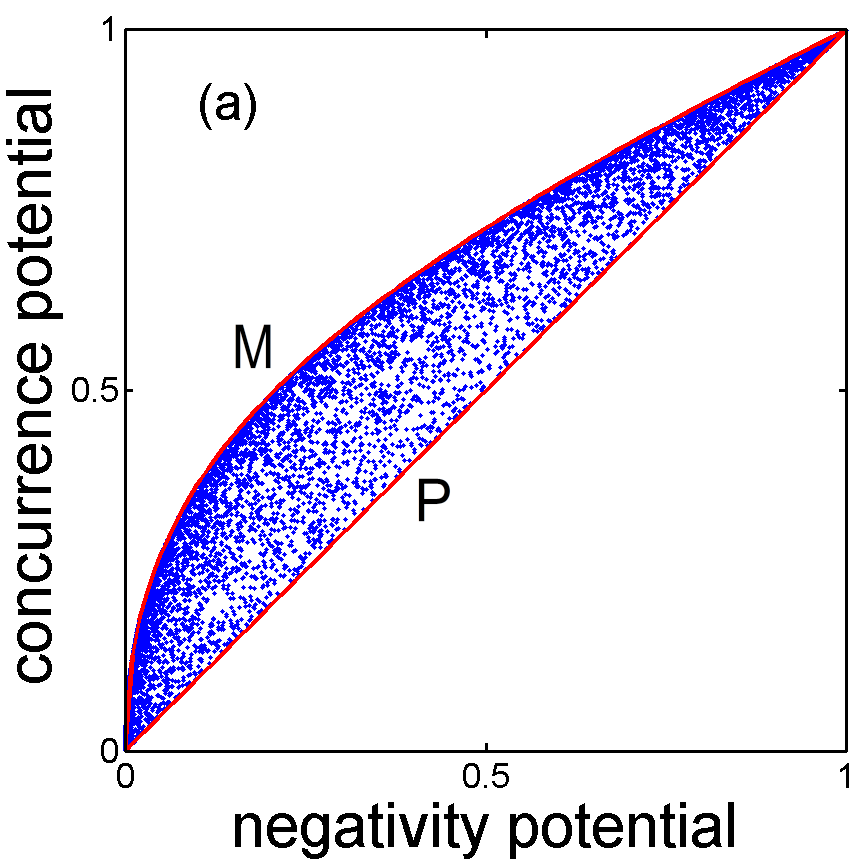}
\includegraphics[width=6cm]{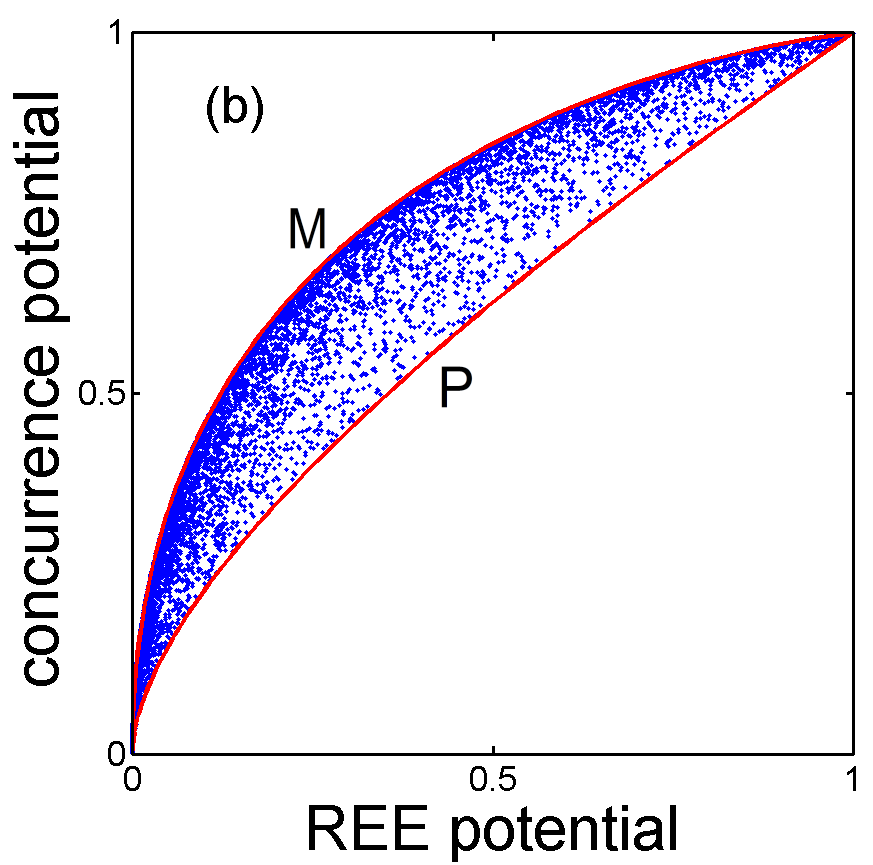}
\includegraphics[width=6cm]{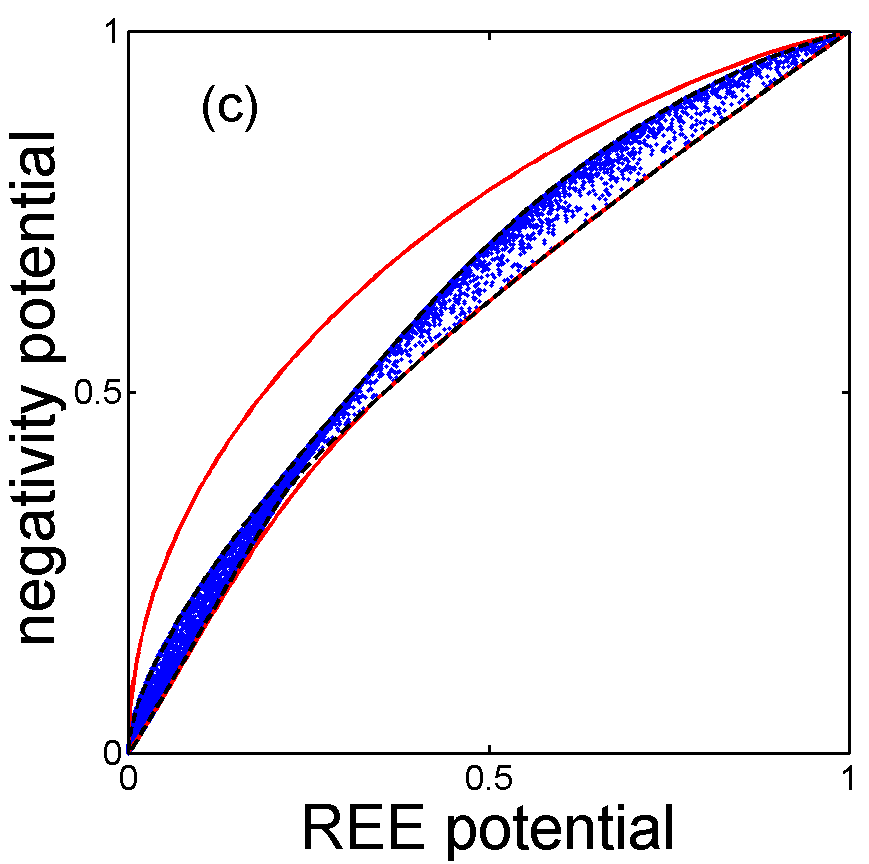}
 \caption{(Color online) Entanglement potentials
for 15,000 single-qubit states $\sigma$ generated via a Monte
Carlo simulation. These entanglement potentials correspond to the
entanglement measures for the two-qubit states $\rho_{\rm out}$
generated from $\sigma$ by a balanced beam splitter. The solid red
curves show the boundaries of the allowed entanglement for
arbitrary two-qubit states $\rho$ (discussed in Appendix~B). Note
that the simulated states $\rho_{\rm out}$ lie in the area bounded
by the solid red curves for the concurrence potential for given
values of both (a) the negativity potential and (b) the potential
of the relative entropy of entanglement (REE). This is not the
case for (c) the negativity potential for a given value of the REE
potential, as the states $\rho_{\rm out}$ lie in the area, bounded
by the dashed black curves, which is smaller than the area covered
by the states $\rho$. The upper (lower) red curves in panels (a)
and (b) correspond to the completely-dephased states $\SIGMA{M}$
(pure states $\SIGMA{P}$) as compactly indicated by M (P). The
meaning of the red and black curves in panel (c) is more detailed,
as shown in Fig.~3.}
\end{figure}
%------------------------------------------------------------------
\begin{figure}
\fig{\includegraphics[width=8cm]{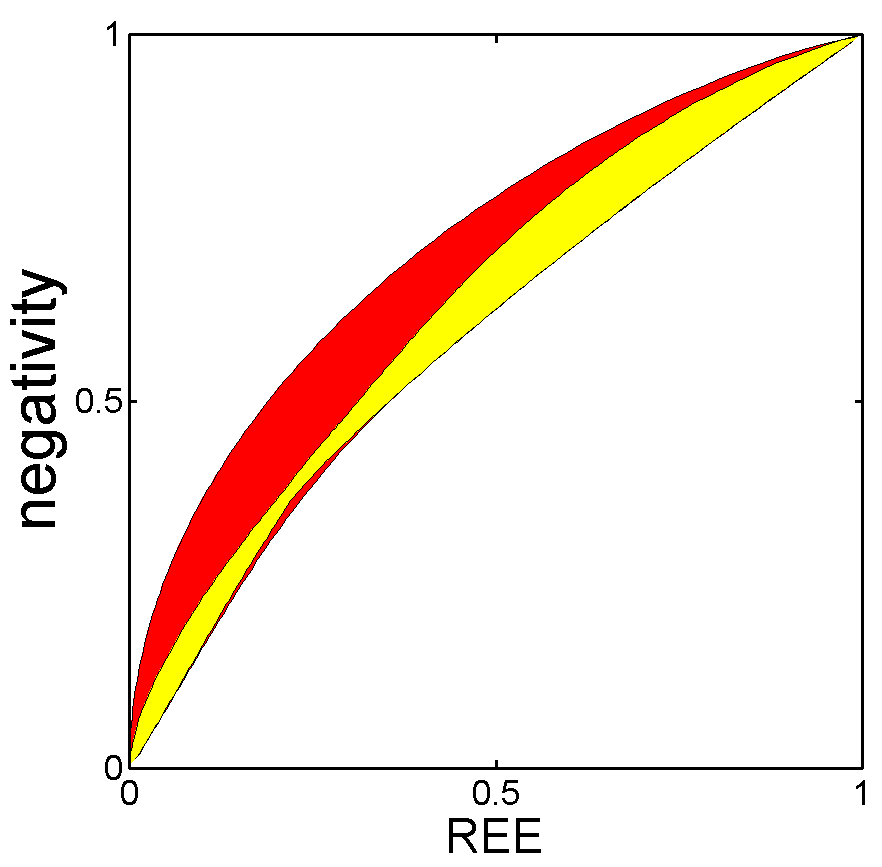}}

\caption{(Color online) Allowed values of the negativity for a
given value of the REE for the states $\rho_{\rm out}$, defined in
Eq.~(\ref{RhoOut1}), generated from arbitrary single-qubit states
(in yellow regions), and those for arbitrary two-qubit states
$\rho$ (in red and yellow regions). The main goal of this paper is
to show how the entanglement, corresponding to any point in the
red regions, can be generated from a single-qubit state.}
\end{figure}
%------------------------------------------------------------------
\subsection{When pure states are maximally nonclassical}

The most general single-qubit pure state is given by
\begin{equation}
  \ket{\psi} = \sqrt{1-p}\ket{0}+e^{i\phi}\sqrt{p}\ket{1},
 \label{psi_p}
\end{equation}
up to an irrelevant global phase. So,
$\SIGMA{P}\equiv\ket{\psi}\bra{\psi}$ is a special case of
Eq.~(\ref{rho}) for $|x|=\sqrt{p(1-p)}$ and $\phi={\rm Arg(}x)$.
As already mentioned, the nonclassicality measures are insensitive
to the relative phase $\phi$, so we can  set $\phi=0$. The
entanglement potentials for a pure state $\SIGMA{P}$ are simply
given by:
\begin{eqnarray}
  \CP(\SIGMA{P}) &=& \NP(\SIGMA{P}) =p,\\
  \REEP(\SIGMA{P}) &=&
  h\left(\textstyle{\frac{1}{2}}[1+\sqrt{1-p^2}]\right),
\label{REEP_pure}
\end{eqnarray}
where $h(y)=-y\log_2 y-(1-y)\log_2(1-y)$ is the binary entropy.
For clarity, we recall the well-known property that
$\REEP(\SIGMA{P})$ is equal to the entanglement of formation
$E_F(\RHO{P})$ and to the von Neumann entropy of  a reduced
density matrix, say, $\tr_1(\RHO{P})$.

We find that pure states are the most nonclassical single-qubit
states in terms of: (i) the maximal NP for a given value of
$\CP\in[0,1]$ [see Fig.~1(a)], (ii) the maximal REEP for a given
value of $\CP\in[0,1]$ [see Fig.~1(b)], and (iii) the maximal REEP
for a given value of $\NP\in[N_2,1]$, where $N_2\approx 0.527$
[see Figs.~1(c), 2 and~3]. These results are summarized in
Table~I. Note that pure states are very close to maximal
nonclassical states concerning the largest NP for a given $\REEP
\lesssim 0.1$ [see Fig.~3].

%\begin{widetext}
%------------------------------------------------------------------
\begin{table}[ht] % TABLE I
\caption{Maximally-nonclassical states (MNS) of a single-qubit
state $\sigma$, maximally-entangled states (MES) $\rho_{\rm out}$
(generated by a lossless balance BS), and maximally-entangled
states $\rho$ (generated by any method) according to one
entanglement measure (or entanglement potential) for a given value
of another entanglement measure (or entanglement potential). The
extra resources  enable to increase the entanglement of $\rho_{\rm
out}$  (so, also the nonclassicality of $\sigma).$ These include
the amplitude-damping channel (ADC), phase-damping channel (PDC),
and tunable beam splitter (TBS). Notation: CP is the concurrence
potential, NP is the negativity potential, REEP is the potential
of the relative entropy of entanglement. The special values of
$N_i$ and $E_i$ are defined in Fig.~3. The output extremal states
read: $\RHO{B}$ are the Bell-diagonal states, $\RHO{P}$ are pure
states, $\RHO{H}$ are the Horodecki states corresponding to the
completely-dephased single-qubit states $\SIGMA{M}$, $\RHO{Z}$ are
the two-qubit states generated from single-qubit
optimally-dephased states $\SIGMA{Z}$, and $\RHO{A}$ are the
optimal generalized Horodecki states. This is a summary of all the
cases, when the nonclassicality of $\sigma$, as quantified by the
three standard entanglement potentials, can or cannot be
increased. }
\begin{tabular}{l l l l l l}
\hline\hline
 Potential 1 & for a given value & MNS & MES & MES & extra resources  \\
 & of Potential 2 & $\sigma$ & $\rho_{\rm out}$ & $\rho$ &
\\ \hline
CP & NP$\in[0,1]$   & $\SIGMA{M}$ & $\RHO{H}$ & $\RHO{H}$ & --- \\
CP & REEP$\in[0,1]$ & $\SIGMA{M}$ & $\RHO{H}$ & $\RHO{H}$ & --- \\
[3pt] %
NP & CP$\in[0,1]$     & $\SIGMA{P}$ & $\RHO{P}$ & $\RHO{P}$ & --- \\
NP & REEP$\in[0,E_3)$ & $\SIGMA{Z}$ & $\RHO{Z}$ & $\RHO{B}$ & PDC \\
NP & REEP$\in[E_3,1]$ & $\SIGMA{M}$ & $\RHO{H}$ & $\RHO{B}$ & PDC \\
[3pt] %
REEP & CP$\in[0,1]$   & $\SIGMA{P}$ & $\RHO{P}$ & $\RHO{P}$ & --- \\
REEP & NP$\in[0,N_1)$ & $\SIGMA{M}$ & $\RHO{H}$ & $\RHO{A}$ & ADC or TBS \\
REEP & NP$\in[N_1,N_2)$&$\SIGMA{P}$ & $\RHO{P}$ & $\RHO{A}$ & ADC or TBS \\
REEP & NP$\in[N_2,1]$  &$\SIGMA{P}$ & $\RHO{P}$ & $\RHO{P}$ & --- \\
\hline\hline
\end{tabular}
\end{table}
%\end{widetext}
%------------------------------------------------------------------
\subsection{When completely-dephased states are maximally  nonclassical}

Now we consider the nonclassicality of a statistical mixture of
the vacuum $\ket{0}$ and single-photon state $\ket{1}$. This is a
special case of Eq.~(\ref{rho}) for the vanishing coherence
parameter $x=0$, i.e.,
\begin{equation}
  \SIGMA{M} = \sigma(p,x=0)= (1-p)\ket{0}\bra{0}+p\ket{1}\bra{1}.
 \label{RhoM}
\end{equation}
These mixtures are referred here to as completely-dephased states,
but can also be referred to as completely-mixed
states~\cite{Miran15}, or the Fock diagonal states, to emphasize
that these states are diagonal in the Fock basis. As shown in
Ref.~\cite{Miran15}, these states can be considered the most
nonclassical by comparing the CP for a given value of the NP.
Here, we analyze the nonclassicality of $\SIGMA{M}$ also with
respect to the REE potential.

First we recall that $\SIGMA{M}$ is transformed by the balanced BS
(with the vacuum in the other port) into the Horodecki state
\begin{equation}
  \RHO{H}(p) = \rho_{\rm out}(p,0) =  p\ket{\psi^-}\bra{\psi^-}+(1-p)\ket{00}\bra{00},
 \label{rhoH}
\end{equation}
which is a mixture of a maximally-entangled state, here the
singlet state $\ket{\psi^-}=(\ket{10}-\ket{01})/\sqrt{2}$, and a
separable state (here, the vacuum) orthogonal to it. The
entanglement properties of the Horodecki state were studied
intensively (see Ref.~\cite{Horodecki09review} for a review), so
we can instantly write the entanglement potentials for $\SIGMA{M}$
as:
\begin{eqnarray}
  \CP(\SIGMA{M}) &=& p, \notag \\
  \NP(\SIGMA{M}) &=& \sqrt{(1-p)^2+p^2}-(1-p), \label{REEP_H} \\
  \REEP(\SIGMA{M}) &=& (p-2)\log_2 (1-p/2)+(1-p)\log_2 (1-p).\notag
\end{eqnarray}
Thus, the REE potential can easily be expressed as a function of
the other potentials $\NP=N$ and $\CP=C$, as follows:
\begin{equation}
  \REEP[\SIGMA{M}(C)]
   = \REEP[\SIGMA{M}(\sqrt{2N(1+N)}-N)].
 \label{REEP_H2}
\end{equation}
We observe that the completely-dephased states $\SIGMA{M}$ are the
most nonclassical single-qubit states concerning: (i) the maximal
CP for a given value of $\NP\in[0,1]$ [see Fig.~1(a)], (ii) the
maximal CP for a given value of $\REEP\in[0,1]$ [see Fig.~1(b)],
(iii) the maximal NP for a given value of the REEP, $E\in[E_3,1]$,
where $E_3=0.397$ [see Fig.~3], and (iv) the maximal REEP for a
given value of $\NP\in[0,N_1]$, where $N_1=0.377$. These results
are also summarized in Table~I.

%------------------------------------------------------------------
\begin{figure}
\fig{\includegraphics[width=8cm]{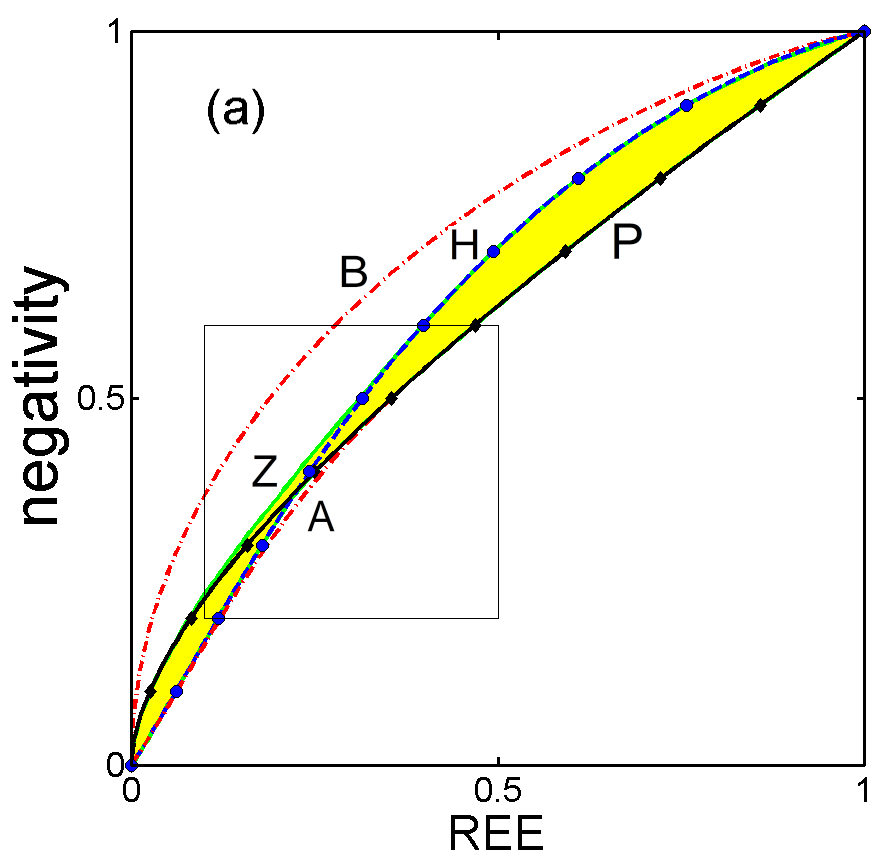}}

\fig{\includegraphics[width=8cm]{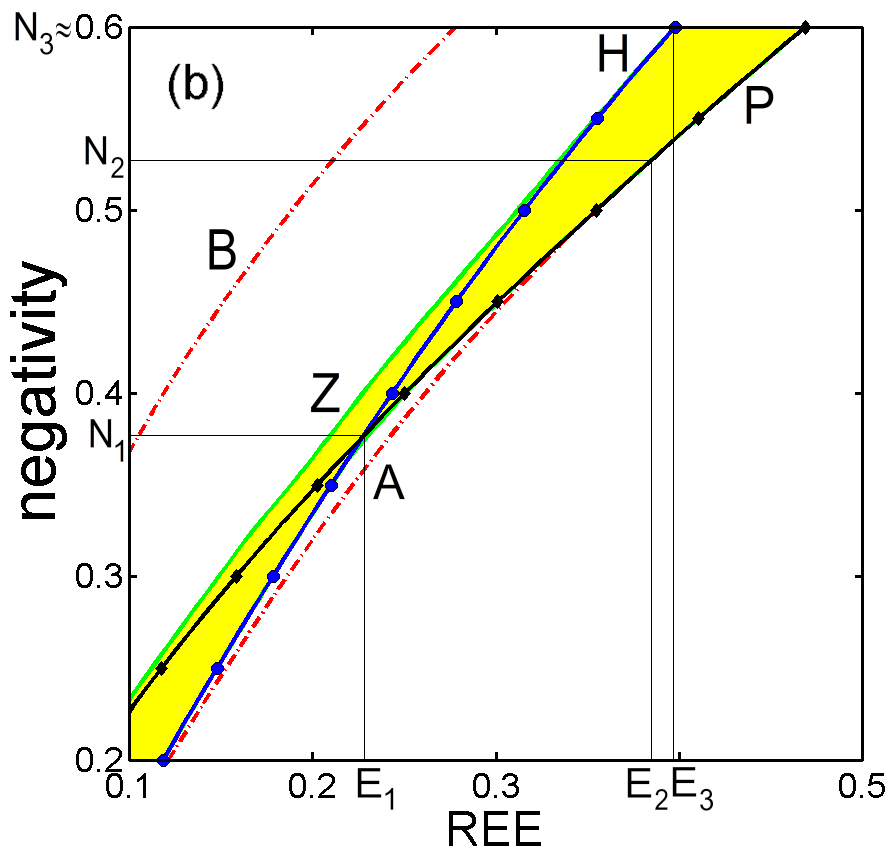}}

\caption{(Color online) (a) Boundary states and special points
corresponding to Fig.~2. (b) The inset of panel (a) showing, in
greater detail, the boundary states: $\RHO{B}$ are the
Bell-diagonal states (corresponding to the upper red dot-dashed
curve), $\RHO{P}$ (or equivalently $\SIGMA{P}$) are pure states
(solid black curve with diamonds), $\RHO{H}$ are the Horodecki
states (solid blue curve with  circles), which correspond to the
single-qubit completely-dephased states $\SIGMA{M}$, $\RHO{A}$ are
the optimal generalized Horodecki states (lower red dot-dashed
curve), and $\RHO{Z}$ are the two-qubit states (green solid curve,
which is the upper bound of the yellow region) generated from
single-qubit optimally-dephased states $\SIGMA{Z}$. Special points
are marked at: $(E_1\approx 0.228,N_1\approx 0.377)$, $(E_2
\approx 0.385,N_2 \approx 0.527)$, and $(E_3 \approx 0.397,N_3
\approx 0.6)$. Note that here, and in Fig.~2, we refer to
entanglement measures rather than directly to the corresponding
entanglement potentials, because some of the marked regions and
curves cannot be reached by the states generated by the standard
entanglement potentials.}
\end{figure}
%------------------------------------------------------------------
\begin{figure}
\includegraphics[width=7cm]{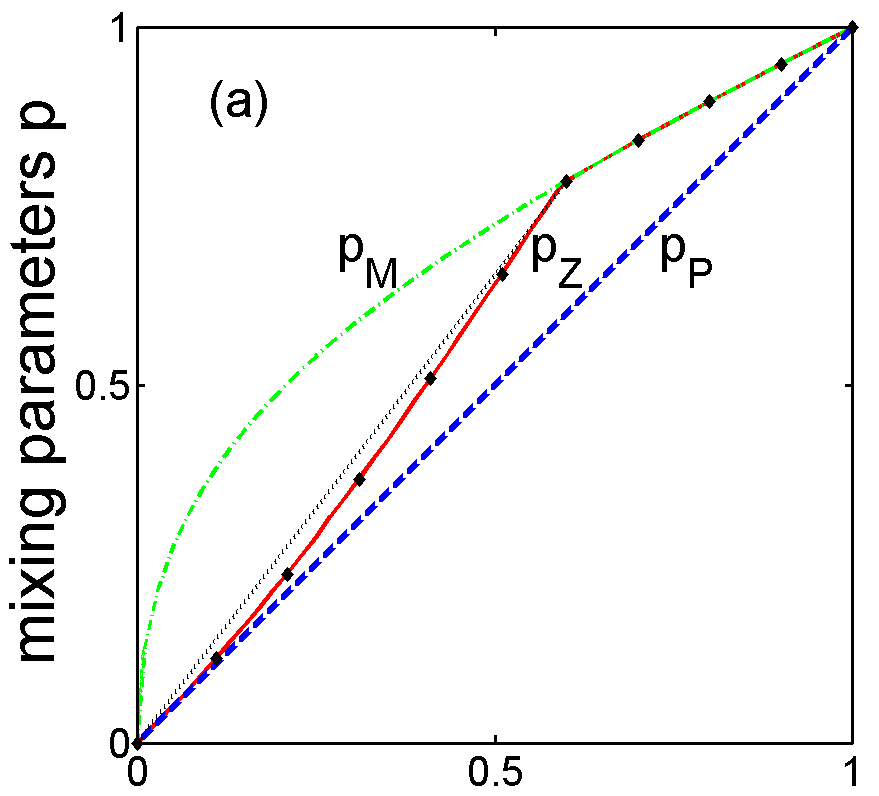}

\includegraphics[width=7cm]{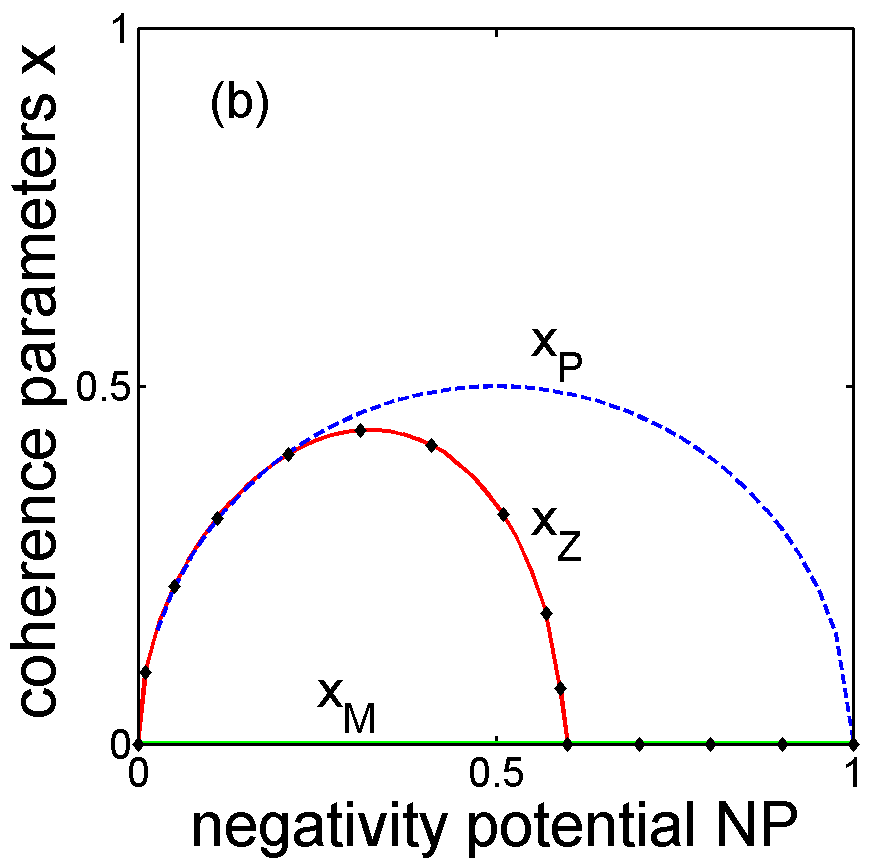}

\caption{(Color online) Mixing and coherence parameters for the
optimally-dephased state $\SIGMA{Z}=\sigma(p_{\rm Z},x_{\rm Z})$,
completely-dephased state $\SIGMA{M}=\sigma(p_{\rm M},x_{\rm
M}=0)$, and pure state $\SIGMA{P}=\sigma(p_{\rm P},x_{\rm P})$ as
a function of their negativity potential
$N=\NP(\SIGMA{Z})=\NP(\SIGMA{M})=\NP(\SIGMA{P})$. Dotted black
line in panel (a) is added, to show that $p_Z$ does not linearly
depend on $N\in[0,0.6]$. We recall that $\SIGMA{Z}$ exhibits the
highest nonclassicality if considered the maximum negativity
potential for a given value of the REEP.}
\end{figure}
%------------------------------------------------------------------
\subsection{When partially-dephased states are maximally
nonclassical}

A closer analysis of Fig.~3 shows that pure states and
completely-dephased states do not have always the greatest
entanglement potentials. Thus, let us define the following
optimally-dephased states
\begin{eqnarray}
  \SIGMA{Z}(N) &=&\sigma[p_{\rm opt},x_{\rm opt}=f(p_{\rm
  opt},N)], \label{rhoX}\\
  \RHO{Z} &\equiv& U_{\rm BS} \big(\SIGMA{Z}\otimes \ket{0}\bra{0} \big)U^\dagger_{\rm
  BS},
\end{eqnarray}
which are the most nonclassical concerning the largest NP for a
given value of the REEP. Here, $f$ is given in Eq.~(\ref{f}),
$U_{\rm BS}$ is defined below Eq.~(\ref{RhoOut1}), while the
optimal mixing parameter $p_{\rm opt}\equiv p_{Z}$ is found
numerically, as $\REEP\{\sigma[p_{\rm opt},f(p_{\rm
opt},N)]\}=\min_p \REEP\{\sigma[p,f(p,N)]\}$, and shown in
Fig.~4(a). Note that the optimal coherence parameter $x_{\rm
opt}\equiv x_Z$, which is shown in Fig.~4(b), is simply given by
$f(p_{\rm opt},N)$. We numerically found that $\SIGMA{Z}$ becomes
$\SIGMA{M}$ for $N\ge N_3\approx 0.6$, as shown in Fig.~3. Thus,
these partially-dephased states become completely dephased.
Moreover, the $\SIGMA{Z}$ are very close to pure states
$\SIGMA{P}$ for $\NP \lesssim 0.2$. The optimal states $\SIGMA{Z}$
are the most distinct from $\SIGMA{P}$ and $\SIGMA{M}$ for $\NP$
near $N_1$ [see Fig.~3(b)].

%------------------------------------------------------------------
\section{Special values of entanglement potentials}

Here we analyze in detail three characteristic points shown in
Fig.~3 corresponding to the negativity (or negativity potential)
for a given value of the REE (or the REE potential).

Point 1: For the negativity potential $N_1\approx 0.377$ and the
REE potential $E_1\approx 0.228$, it holds that pure and
completely-dephased states have the same negativity and REE
potentials, i.e.,
\begin{eqnarray}
  \NP[\SIGMA{P}(N_1)]&=&\NP[\SIGMA{M}(N_1)]=N_1, \notag\\
  \REEP[\SIGMA{P}(N_1)]&=&\REEP[\SIGMA{M}(N_1)]=E_1.
 \label{point1}
\end{eqnarray}
Then one observes that
\begin{eqnarray}
 \NP(\SIGMA{P}) > \NP(\SIGMA{M}) & \quad &{\rm for}\; 0<\REEP<E_1,
 \notag
\\
  \NP(\SIGMA{P}) < \NP(\SIGMA{M}) & \quad &{\rm for}\; E_1<\REEP<1.\quad \label{N09b}
\end{eqnarray}
Point 2: For the negativity potential $N \ge N_2 \approx 0.527$
corresponding to the REE potential $E \ge E_2 \approx 0.385$, one
finds that the optimal generalized Horodecki state $\RHO{A}$, ,
defined in Eq.~(\ref{rhoA}), which maximizes the REE for a given
value of the negativity for arbitrary two-qubit states, becomes a
two-qubit pure state $\RHO{P}$ corresponding to a single-qubit
pure state $\SIGMA{P}$, i.e.,
\begin{eqnarray}
  \NP[\SIGMA{P}(N)]&=&N[\RHO{A}(N)]=N, \notag\\
  \REEP[\SIGMA{P}(N)]&=&E_R[\RHO{A}(N)]=E,\quad \text{if}\;N\ge N_2.\quad
 \label{point2}
\end{eqnarray}
Point 3: For the negativity potential $N \ge N_3 \approx  0.6$ and
the REE potential $E \ge E_3 \approx 0.397$, we find that the
optimally-dephased state $\SIGMA{Z}$, which maximizes the
negativity potential for a given value of the REE potential,
becomes a completely-dephased state $\SIGMA{M}$, i.e.,
\begin{eqnarray}
  \NP[\SIGMA{M}(N)]&=&\NP[\SIGMA{Z}(N)]=N, \notag\\
  \REEP[\SIGMA{M}(N)]&=&\REEP[\SIGMA{Z}(N)]=E,\; \text{if}\;N\ge N_3.\quad\quad
 \label{point3}
\end{eqnarray}
Moreover, although it is not clear on the scale of Fig.~3, the
optimally-dephased state $\SIGMA{Z}$ becomes exactly a pure state
$\SIGMA{P}$ only for the vacuum and single-photon states, i.e.,
\begin{eqnarray}
  \NP[\SIGMA{P}(N)]&=&\NP[\SIGMA{Z}(N)]=N, \notag\\
  \REEP[\SIGMA{P}(N)]&=&\REEP[\SIGMA{Z}(N)]=E,\; \text{if}\;N =0,1.\quad\quad
 \label{point4}
\end{eqnarray}
Analogously, the optimal generalized Horodecki state $\RHO{A}$
becomes exactly the standard Horodecki state $\RHO{H}$, which can
be generated from a completely-dephased state $\SIGMA{M}$, only
for the same cases, as in Eq.~(\ref{point4}), i.e.,
\begin{eqnarray}
   N[\RHO{H}(N)]&=&N[\RHO{A}(N)]=N, \notag\\
   E_R[\RHO{H}(N)]&=&E_R[\RHO{A}(N)]=E,\quad \text{if}\;N =0,1.\quad
 \label{point5}
\end{eqnarray}
Nevertheless, $\RHO{H}$ is a good approximation of $\RHO{A}$, and
$\RHO{P}$ is a good approximation of $\RHO{Z}$ for much larger
ranges of $N$, as shown in Fig.~3.

%------------------------------------------------------------------
\section{Quantifying nonclassicality by generalized entanglement potentials}

Here we address the question of how to generate entanglement, from
single-qubit states $\sigma$, corresponding to the ``forbidden''
red regions shown in Fig.~2. Thus, we define generalized
entanglement potentials, which are the standard entanglement
measures calculated not for the output state $\rho_{\rm out}$,
given in Eq.~(\ref{RhoOut1}) of the non-dissipative balanced BS,
but for the output state $\rho'_{\rm out}$ of a BS, which can be
dissipative and unbalanced. So, can write:
\begin{eqnarray}
  \text{GNP}(\sigma) &=& N(\rho'_{\rm out}), \\
  \text{GCP}(\sigma) &= &C(\rho'_{\rm out}), \\
  \text{GREEP}(\sigma) &= &E_R(\rho'_{\rm out}),
\label{GEP}
\end{eqnarray}
as a generalization of Eqs.~(\ref{NP})--(\ref{REEP}).

%------------------------------------------------------------------
\subsection{How to increase the relative nonclassicality by phase damping}

Here we show that the nonclassicality of a single-qubit state, as
quantified by the negativity for a given value of the REE,
$E_R\in(0,1)$, can be increased by phase damping.

We recall that a phase-damping channel (PDC) for the $i$th  qubit
can be described by the following Kraus
operators~\cite{NielsenBook}:
\begin{equation}
E_{0}(\kappa_{i})=|0\rangle\langle0|+\sqrt{1-\kappa_{i}}|1\rangle\langle1|,\quad
E_{1}(\kappa_{i})=\sqrt{\kappa_{i}}|1\rangle\langle1|,\label{Kraus_pdc}
\end{equation}
where $\kappa_i$ is the phase-damping coefficient and $i=1,2$. Let
us analyze a pure state
\begin{equation}
|\psi_q \rangle =\sqrt{q} |01\rangle + \sqrt{1-q} |10\rangle,
\label{psi_q}
\end{equation}
where $q\in[0,1].$ Note that a general single-qubit pure state,
given in Eq.~(\ref{rho}) for $|x|^2=p(1-p)$, can be simplified by
local rotations to $|\psi_q \rangle$. One can find that a given
pure state $|\psi_q\rangle$ is changed by the PDC into a mixed
state, which can be given  in the Bell-state basis~as
follows~\cite{Horst13}:
\begin{eqnarray}
\RHO{PDC}(q,\kappa_{1},\kappa_{2})=(\textstyle{\frac{1}{2}}-y)|\beta_{1}\rangle\langle\beta_{1}|
+(\textstyle{\frac{1}{2}}+y)|\beta_{2}\rangle\langle\beta_{2}|\nonumber \\
 +(q-\textstyle{\frac{1}{2}})(|\beta_{1}\rangle\langle\beta_{2}|+|\beta_{2}\rangle\langle\beta_{1}|),
\hspace{5mm}
\end{eqnarray}
where $\ket{\beta_{1,2}}=\ket{\psi_{\mp}}$ and
$y=\sqrt{q(1-q)(1-\kappa_{1})(1-\kappa_{2})}$. Now we set $q=1/2$
or, equivalently, we choose the input state to be $\ket{1}$, which
becomes $\rho_{\rm out}(p=1,x=0)$, given by Eq.~(\ref{RhoOut1}).
Then after the PDC transformation, $\rho_{\rm out}$ is changed
into a Bell-diagonal state
\begin{equation}
\RHO{B}=\RHO{PDC}(\tfrac12,\kappa_{1},\kappa_{2})=\lambda_-|\beta_{1}\rangle\langle\beta_{1}|
+\lambda_{+}|\beta_{2}\rangle\langle\beta_{2}|, \hspace{5mm}
\label{rhoB1}
\end{equation}
where $\lambda_{\pm}=[1\pm\sqrt{(1-\kappa_{1})(1-\kappa_{2})}]/2$,
which is a special case of Eq.~(\ref{rhoB}). By applying
Eq.~(\ref{NCErhoB}), one obtains
\begin{eqnarray}
   N(\RHO{B}) &=& C(\RHO{B})=\sqrt{(1-\kappa_{1})(1-\kappa_{2}),}
\label{NCErhoB2}
\end{eqnarray}
which can be changed from zero to one by changing the
phase-damping coefficients. Equation~(\ref{NCErhoB2}) clearly
shows that the PDC can be used for one or both output modes.

We can summarize that it is possible to increase the
nonclassicality of an input state by the phase damping of the BS
output state. Specifically, one can increase the negativity for a
given value of the REE $E_R\in(0,1)$ in comparison to that
predicted by the standard entanglement potentials using a balanced
BS without damping. This increased nonclassicality is shown by the
upper red crescent-shape region in Fig.~2, where the uppermost
curve corresponds to the entanglement of the Bell-diagonal states
$\RHO{B}$. However, it should be stressed that the entanglement
and nonclassicality measures are not increased by the phase
damping channels (since they act locally on the output), but the
possible ratios of different measures can be increased.
Specifically, we start from a highly nonclassical state and
decrease its entanglement (and nonclassicality) via this phase
damping in a such way that the final state has the entanglement,
corresponding to the upper red region in Fig.~2, which cannot be
generated from a single-qubit state using a BS without damping.

%------------------------------------------------------------------
\subsection{How to increase the relative nonclassicality by amplitude damping}

Now we show that the nonclassicality quantified by the REE for a
given value of the negativity, $N\in(0,N_2),$ can be increased by
amplitude damping. Here we assume that the BS is balanced (not
tunable), but we place an amplitude-damping channel in both (or
even single) output modes (ports).

An amplitude-damping channel (ADC) for the $i$th  qubit can be
described by the following Kraus operators~\cite{NielsenBook}:
\begin{equation}
E_{0}(\gamma_{i})=|0\rangle\langle0|+\sqrt{1- \gamma_{i}}
|1\rangle\langle1|,\quad
E_{1}(\gamma_{i})=\sqrt{\gamma_{i}}|0\rangle\langle1|,\label{Kraus_adc}
\end{equation}
where $\gamma_{i}$ is the amplitude-damping coefficient, and
$i=1,2$. As can easily be verified (see, e.g.,
Refs.~\cite{Horst13,Bartkiewicz13}), a pure state
$|\psi_{q}\rangle$, given in Eq.~(\ref{psi_q}), is changed by the
ADC into the mixed state
\begin{eqnarray}
\RHO{ADC}(q,\gamma_{1},\gamma_{2})&=& \RHO{GH}(p',q')\nonumber\\
&=& p' |\psi_{q'}\> \< \psi_{q'}| + (1-p')|00\>\<00|,\quad
\label{rho_adc}
\end{eqnarray}
which is the generalized Horodecki state, given in
Eq.~(\ref{rhoGH}) for $p'=1-(1-q)(1-\gamma_{1})-q(1-\gamma_{2})$
and $q'=q (1-\gamma_{2})/(1-p')]$. Note that $1-p'$ can be
considered an effective damping constant of the pure
$|\psi_{q'}\rangle$, given by Eq.~(\ref{psi_q}) but for $q'$
specified above. Note that by choosing properly the parameters
$q,\gamma_{1},\gamma_{2}$, the amplitude-damped state $\RHO{ADC}$
can be changed into the optimal generalized Horodecki states
$\RHO{A}$, defined in Eq.~(\ref{rhoA}).

Thus, we have shown that the nonclassicality of an input state can
be increased by the amplitude damping of the BS output state in
such a way that the REE for a given value of the negativity,
$N\in(0,N_2)$, is increased in comparison to the maximum
nonclassicality predicted by the standard entanglement potentials.
This extra nonclassicality corresponds to the lower red
crescent-shape region in Fig.~2, where the lower boundary
corresponds to the optimal generalized Horodecki states $\RHO{A}$.

Analogously to the explanation in Sec. V.A, it is important to
clarify that the amplitude damping applied locally cannot increase
``good'' entanglement measures, but it can increase the ratios of
different measures. Thus, by having a highly nonclassical state,
one can decrease its entanglement via amplitude damping in a such
manner that the final damped state has the entanglement
corresponding to some point in the lower red region in Fig.~2,
which cannot be generated in the standard approach, i.e., from a
single-qubit state via a balanced lossless BS.

%------------------------------------------------------------------
\subsection{How to increase the relative nonclassicality by unbalanced beam splitting}

The effect of amplitude damping can be simply modelled by a
tunable lossless BS, as described by $U_{\rm BS}$, given below
Eq.~(\ref{RhoOut1}) but for $\theta\neq \pi/2$. Then, by assuming
that the single-qubit $\sigma$ is given by Eq.~(\ref{rho}), the
two-qubit output state is simply described by
\begin{equation}
  \rho^{\theta}_{\rm out}(p,x) = \left[\begin{array}{cccc}
1-p& -xr & xt& 0\\
-x^*r& pr^2&-prt&0\\
 x^*t& -prt&pt^{2}&0\\
0&0&0&0\\
\end{array}\right],
\label{RhoOutT}
\end{equation}
where $t^2 =T=\cos^2(\theta/2)$ is the BS transmissivity and
$r^2=R=\sin^2(\theta/2)$ is the BS reflectivity. Then, one can
observe that $\rho^{\theta}_{\rm out}$ for $x=0$ reduces to the
generalized Horodecki state $\RHO{GH}$, given in
Eq.~(\ref{rhoGH}), i.e.,
\begin{equation}
  \rho^{\theta}_{\rm out}(p,x=0) = \RHO{GH}(p,q=R=r^2).
 \label{GH_TBS}
\end{equation}
In analogy to the results of Sec. V.B, by choosing properly the
mixing parameter $p$ and the BS reflectivity $R$, one can then
obtain the optimal generalized Horodecki state $\RHO{A}$, defined
by Eq.~(\ref{rhoA}), as a special case of $\RHO{GH}$. The state
$\RHO{A}$ maximizes the REE (or, equivalently, the GREEP) for a
given value of the negativity $N$ (or the GNP) for arbitrary
two-qubit states, as discussed in Appendix~B.

Finally, let us stress again that the optimally-dephased state
$\RHO{A}$ cannot be obtained from a single-qubit state $\sigma$ by
a balanced lossless BS if $N<N_2$ [see Fig.~3]. Thus, such high
nonclassicality of $\sigma$ cannot be measured by applying the
standard entanglement potentials.

These results are summarized in Table~I.

%------------------------------------------------------------------
\section{Conclusions}

This paper addressed the problem of quantifying the
nonclassicality of an arbitrary single-qubit optical state in the
unified picture of nonclassicality and entanglement using the
concept of (standard) entanglement potentials introduced in
Ref.~\cite{Asboth05}. The basis states of the analyzed optical
qubits are the vacuum and single-photon states. In this approach,
the nonclassicality of a single-qubit state is measured by the
entanglement, which can be generated by the light combined with
the vacuum on a balanced lossless beam splitter.

We applied the most popular three measures of two-qubit
entanglement for the states generated with this auxiliary beam
splitter. Specifically, we used the following standard
entanglement potentials of a single-qubit state based on the
negativity (N), concurrence (C), and the relative entropy of
entanglement (REE).

We presented a comparative study of these entanglement potentials
showing a counterintuitive result that an entanglement potential,
for a given value of another entanglement potential, can be
increased by phase and amplitude damping, as well as unbalanced
beam splitting.

The goal of this work was to find the maximal nonclassicality,
corresponding to the maximal value of one entanglement potential
for a fixed value of another entanglement potential, for (i)
arbitrary two-qubit states and (ii) those states which can be
generated from a single-qubit state and the vacuum via a balanced
lossless beam splitter.

We found that the maximal relative nonclassicality measured by the
REE potential for a fixed value (such that $\lesssim 0.527$) of
the negativity potential can be increased by the amplitude damping
of the output state of the balanced beam splitter or,
equivalently, by replacing this beam splitter by a tunable
lossless one. We also showed that the maximal nonclassicality
measured by the negativity potential for a given value (except the
extremal values 0 and 1) of the REE potential can be increased by
phase damping (dephasing). Thus, we introduced the concept of
generalized entanglement potentials in analogy with the standard
potentials, but by allowing unbalanced beam splitting or
dissipation. Of course, the entanglement itself is not increased
by these losses (since they act locally on the output), but the
possible ratios of different measures are affected.

The physical or operational meaning of the standard and
generalized entanglement potentials is closely related to the
corresponding standard entanglement measures. So,  let us recall
the operational meaning of the entanglement measures:

(i) The negativity is a monotonic function of the logarithmic
negativity, which has an operational meaning as a PPT entanglement
cost, i.e., the entanglement cost under the operations preserving
the positivity of the partial transpose
(PPT)~\cite{Audenaert03,Ishizaka04}. The logarithmic negativity is
an upper bound to the entanglement of distillation
$E_D$~\cite{Vidal02}. Note that $E_D$ quantifies the resources
required to extract (i.e., distill) the maximum fraction of the
Bell states from multiple copies of a given partially-entangled
state. The negativity is also a useful estimator of entanglement
dimensionality, i.e., the number of entangled degrees of freedom
of two subsystems~\cite{Eltschka13}. Then, we can interpret both
standard and generalized negativity potentials as the entanglement
potentials for the PPT entanglement cost and for estimating the
entangled dimensions of the light in the beam-splitter outputs.

(ii) The concurrence is monotonically related to the entanglement
of formation, $E_F$~\cite{Wootters98}, which quantifies the
resources required to create a given entangled
state~\cite{Bennett96}. Thus, both standard and generalized
concurrence potentials can also be interpreted as the potentials
for the entanglement of formation. We note that the concurrence
potential of a single-qubit state $\sigma$ can also be
interpreted~\cite{Miran15} as a Hillery-type nonclassical
distance~\cite{Hillery87}, defined by the Bures distance of
$\sigma$ to the vacuum.

(iii) The REE $E_R$ is a convenient geometric measure of the
distinguishability of an entangled state from separable states.
Thus, both standard and generalized REE potentials can be used as
measures of distinguishability of a nonclassical state from
classical states.

Moreover, it is worth noting that the following inequalities
hold~\cite{Horodecki09review}: $$E_F(\rho)\ge E_C(\rho)\ge
E_R(\rho)\ge E_D(\rho)$$ for an arbitrary two-qubit state $\rho$,
where the equalities hold for pure states. Here, $E_C$ is the
(true) entanglement cost. Clearly, the same inequalities hold for
the corresponding entanglement potentials (EP):
$${\rm EP}_F(\sigma)\ge {\rm EP}_C(\sigma)\ge {\rm EP}_R(\sigma)\ge {\rm EP}_D(\sigma),$$
for an arbitrary single-qubit state $\sigma$. Here ${\rm
EP}_F(\sigma)=h(\textstyle{\frac{1}{2}}[1+\sqrt{1-\CP^2(\sigma)}])$
by applying Eq.~(\ref{EoF}). Thus, the REE potential ${\rm
EP}_R(\sigma)\equiv \REEP(\sigma)$ is an upper (lower) bound for
the potential of the entanglement of distillation (formation).
Analogous conclusions can be drawn for the generalized REEP.

Thus, the maximal nonclassicality measured by the standard
negativity potential for a given value of the standard REE
potential can be exceeded (except the values 0 and 1) by the
corresponding generalized potentials. This conclusion can be
rephrased in various ways by recalling multiple physical meanings
of these potentials and the corresponding entanglement measures,
as explained above.

By contrast to these results, we found that the maximal relative
nonclassicality cannot be increased if it is measured by (i) the
concurrence potential for any given value of the negativity
potential, and vice versa, (ii) the concurrence potential for any
fixed value of the REE potential, and vice versa, or (iii) the REE
potential for a fixed value $\gtrsim 0.527$ of the negativity
potential. This is because that, for these three cases, the
generated entanglement in the standard approach of
Ref.~\cite{Asboth05} is exactly the same as the maximal
entanglement of arbitrary two-qubit states.

As discussed in Ref.~\cite{Asboth05}: ``Although we currently lack
of a general proof, all examples we checked analytically and
numerically indicate that the transmissivity of the optimal BS is
1/2 independent of the input state.'' Our examples indicate that
not only the transmissivity $T\neq 1/2$ can lead to higher
nonclassicality, but also adding dissipation increases relative
nonclassicality.

It is worth noting that Refs.~\cite{Horst13,Bartkiewicz13}, which
are closely related to the present study, discussed the effect of
amplitude damping and phase damping on pure states resulting in
increasing the ratios of various measures of entanglement and
Bell's nonlocality. Moreover, Refs.~\cite{Bula13,Meyer13} showed
how to increase the ratios of entanglement measures of
amplitude-damped states by a linear-optical qubit amplifier.

Moreover, it is known that both pure and completely-dephased
single-qubit states can be considered the most nonclassical by
comparing some entanglement potentials~\cite{Miran15}. Here, we
found partially-dephased states, which are the most nonclassical
in terms of the highest negativity potential for a given value of
the REE potential $\lesssim 0.6$.

On the basis of our results, one can infer that some standard
entanglement measures may not be useful for entanglement
potentials. Alternatively, one can conclude that a single balanced
lossless beam splitter is not always transferring the whole
nonclassicality of its input state into the entanglement of its
output modes. The concept of generalized entanglement potentials
can solve this problem at least for the cases analyzed in this
work.

%------------------------------------------------------------------
\acknowledgments

The authors thank Anirban Pathak, Jan Pe\v{r}ina Jr., and Mehmet
Emre Tasgin for stimulating discussions. A.M. gratefully
acknowledges a long-term fellowship from the Japan Society for the
Promotion of Science (JSPS). A.M. is supported by the Grant No.
DEC-2011/03/B/ST2/01903 of the Polish National Science Centre.
K.B. acknowledges the support by the Polish National Science
Centre (Grant No. DEC-2013/11/D/ST2/02638) and by the Ministry of
Education, Youth and Sports of the Czech Republic (Project No.
LO1305). F.N. is partially supported by the RIKEN iTHES Project,
the MURI Center for Dynamic Magneto-Optics via the AFOSR award
number FA9550-14-1-0040, the IMPACT program of JST, and a
Grant-in-Aid for Scientific Research (A).

\appendix

%------------------------------------------------------------------
\section{Definitions of standard entanglement measures}

For the completeness of our presentation, we recall the well-known
definitions of three popular measures of entanglement of two-qubit
states: the negativity, concurrence, and the relative entropy of
entanglement (REE), which are applied in our paper.

(i) The negativity of a bipartite state $\rho$ can be defined
as~\cite{Zyczkowski98,Vidal02}
\begin{equation}
  {N}({\rho})=\max \big[0,-2\min{\rm
eig}(\rho^{\Gamma})\big],
 \label{negativity}
\end{equation}
which is proportional to the minimum negative eigenvalue of the
partially transposed $\rho$, as denoted by $\rho^{\Gamma}$. For
two-qubit states, the minimalization in this definition can be
omitted, as $\rho^{\Gamma}$ can have at most a single negative
eigenvalue. The negativity is a monotonic function of the
logarithmic negativity, $\log_2[N(\rho)+1]$, which can be
interpreted as the entanglement cost under the operations
preserving the positivity of partial transpose~\cite{Audenaert03,
Ishizaka04}. The negativity can also be interpreted as an
estimator of entanglement dimensionality~\cite{Eltschka13}. In
this paper, for simplicity, we apply the potential based on the
negativity instead of  the logarithmic negativity.

(ii) The concurrence for a two-qubit state $\rho$ can be defined
as~\cite{Wootters98}:
\begin{equation}
  C({\rho})=\max \Big\{0,2\lambda_{\max}-\sum_j\lambda_j\Big\},
 \label{concurrence}
\end{equation}
where $\lambda^2 _{j} = \mathrm{eig}[{\rho }(Y\otimes
Y){\rho}^{\ast }(Y\otimes Y)]_j$,
$\lambda_{\max}=\max_j\lambda_j$, $Y$ is the Pauli operator, and
asterisk denotes complex conjugation. The concurrence is
monotonically related to the entanglement of formation,
$E_{F}({\rho})$~\cite{Bennett96} via the binary entropy $h(x)$,
defined below Eq.~(\ref{REEP_pure}), as follows~\cite{Wootters98}:
\begin{equation}
E_{F}({\rho})=
h\left(\textstyle{\frac{1}{2}}[1+\sqrt{1-C^2({\rho})}]\right).
\label{EoF}
\end{equation}
In this paper, we solely apply the concurrence potential rather
than the potential based on the entanglement of formation.

(iii) The REE for a two-qubit state $\rho$ can be defined
as~\cite{Vedral97,Vedral98}:
\begin{equation}
E_R(\rho)=S(\rho ||\rho_{\rm CSS})=\min_{\rho_{\rm sep}\in {\cal
D}} S(\rho ||\rho_{\rm sep}),
 \label{REE}
\end{equation}
which is the relative entropy $S(\rho ||\rho_{\rm sep} )={\rm
Tr}\,( \rho \log_2 \rho -\rho\log_2 \rho_{\rm sep})$ given in
terms of the Kullback-Leibler distance minimized over the set
${\cal D}$ of all two-qubit separable states $\rho_{\rm sep}$.
Thus, $\rho_{\rm CSS}$ denotes a closest separable state (CSS) for
a given $\rho$. Note that the Kullback-Leibler distance is not
symmetric and does not fulfill the triangle inequality, thus it is
not a true metric. The motivation behind using the
Kullback-Leibler distance, instead of other true metrics (like,
e.g., the Bures distance) is the following: For pure states, the
REE based on the Kullback-Leibler distance reduces to the von
Neumann entropy of one of the subsystems. Contrary to the
negativity and concurrence, there is a computational difficulty to
calculate the REE for two-qubit states, except for some special
classes of states. This problem, formulated in
Ref.~\cite{Eisert05}, corresponds to finding an analytical compact
formula for the CSS for a general two-qubit mixed state. As
explained in, e.g., Refs.~\cite{Ishizaka03,Miran08a,Kim10}, this
is very unlikely to solve this problem.  Surprisingly, there is a
compact-form solution of the converse problem: For a given CSS,
all the entangled states (with the same CSS) can be found
analytically not only for two qubits~\cite{Ishizaka03,Miran08a}
but, in general, for arbitrary multipartite systems of any
dimensions~\cite{Friedland11}. As inspired by this approach to the
REE, a general method has been developed recently in
Ref.~\cite{Girard14} to solve the converse problems instead of
finding explicit solutions of convex optimization problems in
quantum information theory.

There are various numerical procedures for calculating the
two-qubit REE~\cite{Vedral98,Miran08b,Zinchenko10,Girard15}.
Probably, the most reliable and efficient is the algorithm of
Ref.~\cite{Girard15} based on semidefinite programing in
CVX~\cite{CVX} (a MATLAB-based modeling system for convex
optimization), which is also applied in this paper.

All these measures vanish for separable states and are equal to
one for the two-qubit Bell states.

%------------------------------------------------------------------
\section{Boundary states for arbitrary two-qubit states}

Here, we recall well-known
results~\cite{Eisert99,Verstraete01,Miran04a,Miran04b,Miran08b,Horst13}
on the boundary states of one entanglement measure for a given
value of another entanglement measure for arbitrary two-qubit
states $\rho$. Note that the two-qubit states $\rho_{\rm out}$,
which can be generated from single-qubit states $\sigma$ and the
vacuum by a balanced BS, are only a subset of the states $\rho$.
These boundary states are shown by red solid curves in Fig.~1.

%------------------------------------------------------------------
\subsection{Pure states}

A two-qubit pure state, $\ket{\psi}=\sum_{n,m=0,1}c_{nm}\ket{nm}$,
including the state
\begin{equation}
|\psi_{\rm out} \rangle
=\sqrt{1-p}|00\>+\sqrt{\tfrac{p}{2}}(|10\>-|01\>),
\label{PsiRhoOut}
\end{equation}
which is generated by a balanced BS from a general single-qubit
pure state can be simplified by local rotations to $|\psi_q
\rangle$, given by Eq.~(\ref{psi_q}). In the case of the state
given by Eq.~(\ref{PsiRhoOut}), it holds $p=2\sqrt{q(1-q)}$. The
negativity and concurrence for $|\psi_q \rangle$ read
\begin{equation}
  N\equiv N(|\psi _{q}\rangle) =C(|\psi _{q}\rangle)
  =2\sqrt{q(1-q)},
 \label{N1}
\end{equation}
and the REE can be given as a function of the negativity (or
concurrence) as follows
\begin{equation}
E_R(|{\psi}_{q}\rangle) = h\left(
\tfrac{1}{2}[1+\sqrt{1-N^{2}}]\right), \label{N05}
\end{equation}
via the binary entropy $h$.

Pure states have the highest entanglement for arbitrary two-qubit
states in the following cases: (i) the maximal negativity for a
given value of the concurrence $C\in[0,1]$ [see Fig. 1(a)], as
shown in Ref.~\cite{Verstraete01}, (ii) the maximal REE for a
given value of the concurrence $C\in[0,1]$ [see Fig. 1(b)], as
shown in Ref.~\cite{Miran04b}, and (iii) the maximal REE for a
given value of the negativity $N\in[N_2,1]$, where $N_2\approx
0.527$ [see Fig. 1(c)] as demonstrated in Ref.~\cite{Miran08b}.

%------------------------------------------------------------------
\subsection{Horodecki states}

The Horodecki states, which are defined in Eq.~(\ref{rhoH}), can
be generated by the balanced BS transformation. These states have
the highest entanglement for arbitrary two-qubit states by
considering: (i) the maximal concurrence for a given value of the
negativity $N\in[0,1]$ [see Fig. 1(a)], as shown in
Ref.~\cite{Verstraete01}, and (ii) the maximal concurrence for a
given value of the REE $E_R\in[0,1]$ [see Fig. 1(b)], as shown in
Ref.~\cite{Miran04b}, and they are (iii) very close to maximal REE
for a given value of the negativity $N\lesssim 0.2$ [see Fig.
1(c)], which was discussed in Ref.~\cite{Miran08b}.

In addition to these two classes of states there are also two
other classes of boundary states, which cannot be generated by a
lossless balanced BS, as discussed below.

%------------------------------------------------------------------
\subsection{Generalized Horodecki states}

A generalized Horodecki state $\RHO{GH}$ can be defined as a
statistical mixture of a pure state $|\psi_q\>$, given by
Eq.~(\ref{psi_q}), and a separable state (say the vacuum)
orthogonal to it, i.e.,~\cite{Miran08a}:
\begin{eqnarray}
  \RHO{GH}(p,q) &=& p |\psi_q\> \< \psi_q| + (1-p)|00\>\<00|,
\label{rhoGH}
\end{eqnarray}
where $p,q\in [0,1]$. When the pure state $|\psi_q\>$ is a Bell
state, then $\RHO{GH}$ becomes the standard Horodecki state, given
in Eq.~(\ref{rhoH}). The negativity and concurrence are simply
given by:
\begin{eqnarray}
N(\RHO{GH}) &=& \sqrt{(1-p)^2+4p^2q(1-q)}-(1-p), \label{N20a} \quad \\
  C(\RHO{GH}) &=& 2p\sqrt{q(1-q)}.
\label{N20b}
\end{eqnarray}
Unfortunately, no compact-form analytical expression for the REE
for the general state $\RHO{GH}$ is known. We can express the
parameter $q$ as a function $f_1(p,N)$ [$f_2(p,C)$] of the mixing
parameter $p$ and the negativity (concurrence) as follows:
\begin{eqnarray}
  q &=& f_1(p,N) = \frac{1}{2p}\left[p\pm \sqrt{p^2-N^2-2N(1-p)}\right]
  \notag \\
  &=& f_2(p,C) = \frac12\Big(1\pm \sqrt{1-(C/p)^2}\Big),
\label{N21}
\end{eqnarray}
by simply inverting formulas in Eqs.~(\ref{N20a})
and~(\ref{N20b}). Thus, one can have an explicit formula for
$\RHO{GH}$ as a function of $N$ and $C$ as follows:
$\RHO{GH}[p,q=f_1(p,N)]$ for $p\ge \sqrt{2N(1+N)}-N$, and
$\RHO{GH}[p,q=f_2(p,C)]$ for $p\ge C$.

The optimal generalized Horodecki state $\RHO{A}$, as shown in
Fig.~3, can be defined as the generalized Horodecki state
$\RHO{GH}$, which maximizes the REE for a given
$N$~\cite{Miran08a}:
\begin{equation}
  \RHO{A}(N)=\RHO{GH}[\bar p_{\rm opt},f_1(\bar p_{\rm opt},N)],
 \label{rhoA}
\end{equation}
where the optimal mixing parameter $\bar p_{\rm opt}(N)$ is chosen
such that
\begin{equation}
  E_R\{\RHO{GH}[\bar p_{\rm opt},f_1(\bar p_{\rm opt},N)]\}\ =\max_p E_R\{\RHO{GH}(p,f_1(p,N)]\}.
 \label{p_opt}
\end{equation}

%------------------------------------------------------------------
\subsection{Bell-diagonal states}

A general Bell-diagonal state is defined by
\begin{eqnarray}
\RHO{B'}=\sum_{i=1}^4 \lambda_i |\beta_{i} \rangle\langle
\beta_{i}|, \label{rhoB}
\end{eqnarray}
which is a statistical mixture of the Bell states $|\beta_{i}
\rangle$, with the normalized weights $\lambda_i$, i.e.,
$\sum_j\lambda_j=1$. The entanglement measures for $\RHO{B'}$ are
given as follows
\begin{eqnarray}
   N(\RHO{B'}) &=& C(\RHO{B'})=2\max(0,\Lambda-1/2)\ \equiv N, \notag \\
  E_R(\RHO{B'})&=& 1-h[(1+N)/2],
\label{NCErhoB}
\end{eqnarray}
where $\Lambda=\max_j\lambda_j$. A well-studied example of the
Bell-diagonal states is the Werner state~\cite{Werner89}:
\begin{equation}
\RHO{W}=\frac{1+2N}{3}|\psi^{-}\rangle \langle
\psi^{-}|+\frac{1-N}{6} I,  \label{N15}
\end{equation}
where $I$ is the two-qubit identity operator and $|\psi^{-}\rangle$
is the singlet state.

The Bell-diagonal states exhibit the highest  negativity for a
given value of the REE, as discussed in Ref.~\cite{Miran04b} and
shown by the red uppermost curve in Figs.~1(c) and~3. Note that
these states, together with pure states, have the highest
negativity for a given value of the concurrence, as discussed in
Ref.~\cite{Verstraete01} and shown by the lowest red line in
Fig.~1(a)]. It is important to mention that the Bell-diagonal
states cannot be generated by a lossless balanced BS.

%------------------------------------------------------------------

\end{document}